\documentclass[aps,preprintnumbers,amsmath,twocolumn,groupedaddress,superscriptaddress,notitlepage,nofootinbib,prl]{revtex4-1}
\usepackage[utf8]{inputenc}
\usepackage{indentfirst}
\usepackage{amssymb}
\usepackage{amsmath,latexsym}
\usepackage{graphicx}
\usepackage{float}
\usepackage[margin=1.0in]{geometry}
\usepackage[mathscr]{euscript}
\usepackage{color}
\usepackage[dvipsnames]{xcolor}
\usepackage{soul}
\usepackage{slashed}
\usepackage[mathscr]{euscript}
\usepackage{feynmf}
\usepackage{subcaption}
\usepackage{braket}

\DeclareMathOperator{\tr}{tr}

\newcommand{\beq}{\begin{equation}}
\newcommand{\eeq}{\end{equation}}
\newcommand{\bea}{\begin{eqnarray}}
\newcommand{\eea}{\end{eqnarray}}

\long\def\beqs#1\eeqs{\beq\begin{split} #1 \end{split}\eeq}
\newcommand{\hgamma}{ \hat{\gamma} }
\newcommand{\hmu}{ \hat{\mu} }
\newcommand{\hpsi}{ \hat{\psi} }
\newcommand{\hg}{ \hat{g} }
\newcommand{\bbM}{\mathbb{M}}

\newcommand{\transpose}{^{\top}}

\definecolor{MyRed}{RGB}{153,0,13}

\usepackage[colorlinks=true,backref=true, linktocpage=true,
citecolor=MyRed,urlcolor=MyRed,linkcolor=MyRed,pdfpagemode=UseOutlines]{hyperref}

\hypersetup{%
  bookmarksnumbered=true,
  pdftitle = {},
  pdfsubject = {},
  pdfauthor = {},
  pdfkeywords = {}
}

\usepackage{microtype}

\usepackage{pgfplotstable,array,siunitx}
\usepackage{multirow}
\usepackage{booktabs}

\usepackage{tikz}
\newcommand{\ditto}{%
    \tikz{
        \draw [line width=0.12ex] (-0.4ex,0) -- +(0.4ex,0.8ex)
            (0.0ex,0) -- +(0.4ex,0.8ex);
        \draw [line width=0.08ex] (-0.6ex,0.4ex) -- +(-0.7em,0)
            (0.6ex,0.4ex) -- +(0.7em,0);
    }%
}

\preprint{INT-PUB-20-030}

\def\MeV{{\,\rm MeV}}
\def\fm{{\,\rm fm}}
\def\av#1{ \left\langle #1 \right\rangle }

\newcommand{\nn}{\nonumber}
\newcommand{\eq}[1]{Eq.~(\ref{#1})}
\newcommand{\fig}[1]{Fig.~\ref{#1}}

\graphicspath{{figures/}}

\begin{document}
\title{Structure Factors of Neutron Matter at Finite Temperature}

\author{Andrei Alexandru}
\email{aalexan@gwu.edu}
\affiliation{Department of Physics,
The George Washington University, Washington, DC  20052}

\author{Paulo Bedaque}
\email{bedaque@umd.edu}
\affiliation{Department of Physics,
University of Maryland, College Park, MD 20742}

\author{Evan Berkowitz}
\email{evanb@umd.edu}
\affiliation{Department of Physics,
University of Maryland, College Park, MD 20742}

\author{Neill C. Warrington}
\email{ncwarrin@uw.edu}
\affiliation{Institute for Nuclear Theory, University of Washington, Seattle, WA 98195-1550}

\preprint{INT-PUB-20-030}

\date{July 31, 2020}

\begin{abstract}
We compute continuum and infinite volume limit extrapolations of the structure factors of neutron matter at finite temperature and density. Using a lattice formulation of leading-order pionless effective field theory, we compute the momentum dependence of the structure factors at finite temperature and at densities beyond the reach of the virial expansion. The Tan contact parameter is computed and the result agrees with the high momentum tail of the vector structure factor. All errors, statistical and systematic, are controlled for. This calculation is a first step towards a model-independent understanding of the linear response of neutron matter at finite temperature, a realm until now little explored.


\end{abstract}

\maketitle

\section{Introduction}
As much as $99\%$  of the gravitational binding energy released in core-collapse supernovae
escapes the star in the form of neutrinos. This enormous flux, when it interacts with the nuclear matter on its way out of the star, is believed to 
be an essential ingredient in the explosion of the star. 
%
Though neutral-current neutrino-neutron scattering is well-described in vacuum by tree level $Z^0$ exchange, neutrino scattering in supernova material is complicated by many-body dynamics induced by the strong force. Due to its non-perturbative nature, however, these effects are hard to calculate (for a review see Ref.~\cite{Burrows:2004vq}). 
%
We compute here the exact structure factors of a spin-balanced neutron gas at leading order in the pionless effective field theory~\cite{doi:10.1146/annurev.nucl.52.050102.090637} using Monte Carlo methods at fugacities of $z\equiv\exp(\beta\mu)=1.0$ and $1.5$. 
The main accomplishments presented here are successful continuum and infinite volume limit extrapolations at such high fugacities. This was made possible by the import of several methods from lattice QCD for the simulation of fermions including: iterative methods, pseudofermions and chronological inverters~\cite{BROWER1997353}.  Since lattice artifacts typically decrease with density, we expect the methods presented here to provide complete control over this system for any fugacity $z\leq 1.5$. Such exact calculations, computed over a range of densities and temperatures, may place the nuclear physics inputs to supernovae simulations on firmer theoretical footing.

The differential cross section of low energy neutrinos off a gas of non-relativistic neutrons is approximately determined by the {\it static} vector and axial structure factors
\beqs
S_V(q) &= \int  d^3\mathbf{r}\ e^{-i \mathbf{q}. \mathbf{r}} 
\braket{ \delta n(0,\mathbf{r}),   \delta n(0,\mathbf{0})} \\
S_A(q) &= \int d^3\mathbf{r}\ e^{-i\mathbf{q}. \mathbf{r}} 
\braket{ \delta S_z(t,\mathbf{r}),   \delta S_z(0,\mathbf{0})} , 
\eeqs 
where $\delta n = n - \braket{n}$ and  $\delta S_z = S_z - \braket{S_z}$ are the fluctuations of the density and spin  \cite{Burrows:2004vq,HOROWITZ2006326}. These quantities are the main object of our calculations. We will also compute the contact $C$, which determines the asymptotic behavior of the particle distribution function.  As we will see below, our result for the contact will serve as a non-trivial check of our structure factors. 

The interaction between neutrons will be described in this paper with the help of pionless effective field theory~\cite{doi:10.1146/annurev.nucl.52.050102.090637,vanKolck:1998bw,Kaplan:1998tg}. Its applicability is restricted to kinematical regimes where momentum transfers between nucleons is below the pion mass, which is roughly satisfied at the temperatures of interest ($T\alt 10$ MeV). The systematic use of pionless effective field theory provides an expansion of observables in powers of typical momentum scales over the pion mass. In this work, we will restrict ourselves to leading order in the low energy expansion, although higher order calculations will certainly be welcome. At this order, the (continuum) hamiltonian is given by
\beq
H = \int{d^3x~\left[
\frac{\nabla \psi^{\dagger} . \nabla \psi }{2M} - g \big(\psi_1^{\dagger}\psi_1\big) \big(\psi_2^{\dagger}\psi_2 \big)  
\right]},
\label{eq:hamiltonian}
\eeq where $\psi$ is a spin doublet of quantized fields destroying a neutron, the index $\sigma=1,2$ distinguishes the two spin components. The coupling constant $g$ is determined by the s-wave scattering length between neutrons.

%

\section{Formalism}
In order to use numerical methods (and to properly define the contact interaction in the hamiltonian in \eq{eq:hamiltonian}) we will use a spatial cubic lattice with spacing $\Delta x$. The lattice hamiltonian  is
\beqs
&H  = \underbrace{\sum_{x x' }\psi^{\dagger}_{x } k_{x x'} \psi_{x' }}_K \underbrace{- \frac{g}{\Delta x^3}\sum_{x}{\big( \psi^{\dagger}_{x1} \psi_{x1}\big)\big( \psi^{\dagger}_{x2} \psi_{x2}\big)}}_{V}, \\
&k_{xx'}  = \sum_{p}{ \frac{p^2}{2M \Delta x^2}e^{-i p\cdot(x-x')}}\,.
\label{eq:lattice-hamiltonian}
\eeqs
where $p$ are lattice momenta with components \hbox{$p_i  = \frac{2\pi}{N_x} n_i$}, and $-\frac{N_x-1}{2}\leq n_i \leq \frac{N_x-1}{2}$.
We leverage a cubic lattice with $N_x^3$ sites. The number and spin operators are $N = \sum_{x}{\psi^{\dagger}_{x}\psi_{x}} $ and  $\mathbf{S} = \sum_{x }{\psi^{\dagger}_{x} \mathbf{\sigma}\psi_{x}} $, and the chemical potentials coupled to each will be denoted $\mu$ and $h$, respectively. The partition function can be written, with the help of the Trotter formula and the Hubbard-Stratanovich transformation, by using standard steps:
\bea
Z  &=& \tr e^{-\beta (H-\mu N)} \approx 
 \tr \prod_{t=1}^{N_t} e^{-\Delta t K}e^{-\Delta t (V-\mu N)}  \nn\\
   &=&  \int{\prod_{x,t} D\hat \psi_{xt}^{\dagger} D\hat \psi_{xt} D A_{xt} \, e^{-S(\hpsi^{\dagger},\hpsi,A)}}~,
\label{eq:three-equal}
\eea where the error involved in \eq{eq:three-equal} is of order ${\cal O}(\Delta t^2)$ with $\Delta t =  \beta/N_t$, and the action $S=S_F+S_A$ is given by the fermionic and auxiliary-field contributions

\bea
S_F &=& - \sum_{xt}{\hpsi^{\dagger}_{xt+1} e^{A_{xt}+\hmu}}\hpsi_{xt}
+ \sum_{x,x',t}{\hpsi_{xt} B_{x x'} \hpsi_{x't}},  \nn\\
 S_A &=& \frac{1}{\hg} \sum_{x,t}{\Big( \text{cosh}(A_{xt})-1\Big)} .
\eea 
The matrix $B_{x x'}$ is the $N_x^3 \times N_x^3$ matrix representing spatial hopping:
\beq
B_{x x'} = \frac{1}{N_x^3}\sum_p{e^{-i p\cdot(x-x')}e^{\hat \gamma \frac{p^2}{2}}} 
\eeq and the parameters of the lattice action are given by
\begin{align}
\mu \Delta t  & = \hmu + \text{log} \frac{f_1(\hg)}{f_0(\hg)}, 
\quad \frac{\Delta t}{ M \Delta x^2}  = \hgamma ~,\nn\\
\frac{g \Delta t}{\Delta x^3}  &= \text{log} \frac{f_2(\hg) f_0(\hg)}{f_1(\hg)^2}, 
\label{eq:ham-to-axn}
\end{align}
where $f_{\alpha}(\hg) \equiv \int_{-\infty}^\infty{dA ~e^{-\frac{\text{cosh}(A)-1}{\hg}}e^{\alpha A}}$. \eq{eq:three-equal} and the mappings in \eq{eq:ham-to-axn} are derived in Ref.~\cite{PhysRevC.101.045805}.

At leading order in the pionless effective theory, the s-wave phase shift is given by $k\cot \delta(k) = -1/a$, with scattering length $a=-18.9$ fm (higher orders give an effective range term, and so on). Thus the coupling constant $g$ is adjusted in order for the theory, in the continuum, infinite volume, zero temperature and $\mu=0$ limits, to reproduce this scattering amplitude.

The way the continuum limit of our lattice theory is approached is subtle. Numerical results indicate that there are terms proportional to powers of $\Delta t/\Delta x^2$ appearing in several quantities. For that reason, we choose to take the ``hamiltonian" limit, where $\Delta t \rightarrow 0$ is taken before  $\Delta x \rightarrow 0$.  In practice, this is accomplished by keeping $\hat\gamma=\Delta t/M\Delta x^2 \ll 1$ as $\Delta x$ is reduced. In the hamiltonian limit we find that
\beq
\frac{1}{Mg} = \frac{c_1}{\Delta x} - \frac{1}{4\pi a},
\label{eq:coupling-tuning}
\eeq
where $c_1^{-1}=5.14435...\,$.
By fixing $M=938\MeV$, the chemical potential $\mu$, the inverse temperature $\beta$, the box size $L$, and the number of spatial and temporal discretization steps $N_x$ and $N_t$ we can use \eq{eq:ham-to-axn} and \eq{eq:coupling-tuning} to compute $\hat g,\, \hat\gamma,\, \hat\mu$ with $\Delta x = L/N_x$ and $\Delta t = \beta/N_t$, and check $\hat\gamma\ll1$.

\section{Methods}
\label{sec:methods}

To sample the grand canonical ensemble in \eq{eq:three-equal} we rewrite the partition function
\beqs
Z &= \int D{\hat\psi_{xt}^\dagger} D{\hat\psi_{xt}} DA_{xt} \, e^{-S_A(A)-\hat\psi^\dagger \bbM(A) \hat\psi}\\
&=\int D\phi_{xt}^\dagger D\phi_{xt} DA _{xt}\, e^{-S_A(A) -\phi^\dagger {\cal M}^{-1}\phi} \,,
\eeqs
using a complex scalar {\em pseudofermion} field $\phi_{xt}$~(no spinor index).
The fermion matrix $\bbM$ is diagonal in spin and in the spin-balanced case it splits into two identical  blocks $\bbM_1=\bbM_2$.
Furthermore,   ${\cal M}= \bbM_1 \bbM_1\transpose$ so $\det \bbM=\det \bbM_1 \bbM_1\transpose = \det {\cal M}$.
The integrand is then positive definite, so we can apply the usual Monte Carlo methods to sample the partition function.
We use Hybrid Monte Carlo~\cite{Duane:1987de} to sample the field $A$: we interleave the sampling of $\phi$ with probability $P(\phi)\propto \exp(-\phi^\dagger {\cal M}^{-1}\phi)$, a modified Gaussian distribution, with updates of $A$ generated by a classical mechanics evolution according to a Hamiltonian $H(\pi,A) = \pi\transpose \pi/2 + V(A)$ where $V(A)=S_A(A) +  \phi^\dagger {\cal M}^{-1}\phi$ and $\pi$ is canonical momentum conjugate to $A$  sampled randomly according to $P(\pi) \propto \exp(-\pi\transpose\pi/2)$ at the beginning of each classical trajectory.

One ingredient required in the classical evolution of $A$ is the evaluation of the derivative $dV/dA$ (the {\em force} term), which involves the calculation of ${\cal M}^{-1}\phi$.
Since the matrix ${\cal M}$ is hermitian positive definite, we use conjugate gradient, an iterative method, to compute ${\cal M}^{-1}\phi$.
The multiplication with $\bbM_1$ can be split in $N_t$ multiplications with diagonal matrix $e^{A+\hat\mu}$ and $N_t$ multiplications with the hopping matrix $B$.
The most time consuming piece is the multiplication by $B$, but this can be done efficiently in
the momentum space where $B$ is diagonal.
Using the fast Fourier transform, the complexity of multiplying with $B$ is reduced to $V_s \log V_s$, with $V_s=N_x^3$ the number of points in a time slice.
The overall complexity of multiplying with ${\cal M}$ is then ${\cal O}(N_t V_s \log V_s)$, much better than the ${\cal O}(N_t V_s^3)$ complexity of the Hybrid Monte Carlo algorithm without pseudofermions we used in our previous study~\cite{Alexandru:2019gmp}.
For our simulations we find that for $N_x\gtrsim 8$ the pseudofermion method wins out.
Another advantage is that this method can be parallelized efficiently by dividing the lattice evenly over the temporal direction.

\section{Results}

\begin{table}[b!]


\begin{tabular*}{0.98\columnwidth}{@{\kern0pt}@{\extracolsep{\stretch{1}}}*{7}{c}@{\kern5pt}}
\toprule
\multicolumn{7}{c}{Physical parameters for our simulations} \\
\rule{0pt}{3ex} $T[\MeV]$ & $V[\fm^3]$ && $z$ && $n \, [\text{fm}^{-3}]$ & $\epsilon_F \, [\text{MeV}]$\\
\midrule
4.14 & $18^3$ && 1.0 && $5.1(2)\times 10^{-3}$ & $5.9(2)$ \\
{\ditto} & \ditto && 1.5 && $8.2(2)\times 10^{-3}$ & $8.0(1)$ \\
\bottomrule
\end{tabular*}

\caption{The temperature, volume, and fugacity are exact inputs to the calculations and have no uncertainty. In contrast, the density $n$, and Fermi energy $\epsilon_F$
are computed quantities and the error bars are obtained by continuum limit extrapolations. }
\label{table:params}
\end{table}

%
%

\begin{figure}[t!]
\includegraphics[width=0.95\columnwidth,trim=0.0cm 0.0cm 1.3cm 1.5cm, clip]{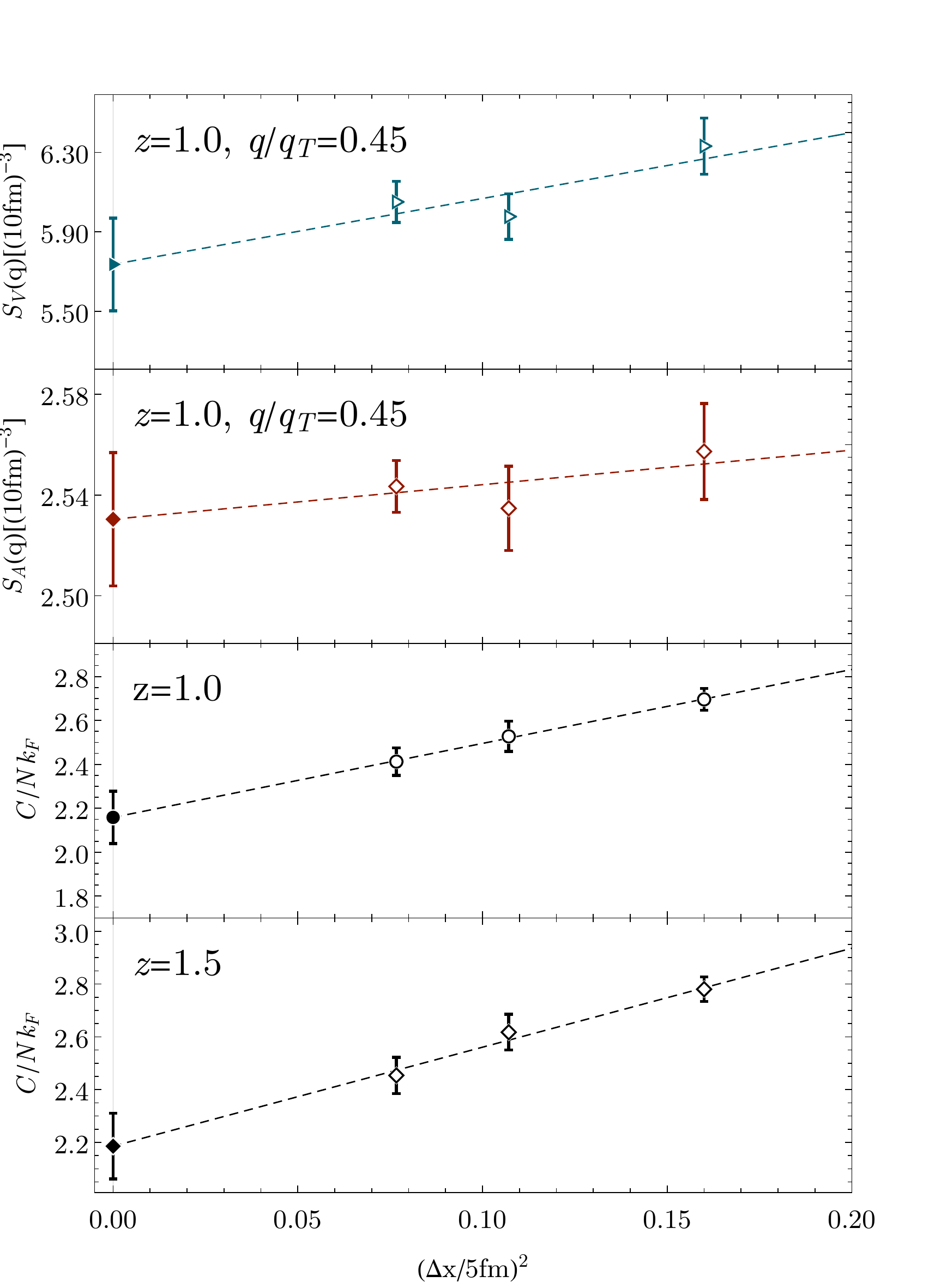}
\caption{Continuum limit extrapolation: for $S_V$ and $S_A$ for $z=1.0$ and $q/q_T = 0.45$~(top) and contact~(bottom).}
\label{fig:dx-cont-lim}
\end{figure}

In all calculations the coupling $g$ was determined with from \eq{eq:coupling-tuning}. As an additional check, we ran simulations with $0.05<z<0.8$, deep in the virial regime, and compared our results with the virial expansion. The second virial coefficient we extracted, $b_2 = 0.419 (3)$, agrees with the Beth-Uhlenbeck prediction $b_2 \simeq 0.415$ \cite{BETH1937915}. Incidental to this check, we estimate the third viral coefficient of this leading order pionless EFT to be $b_3 = -0.13(5)$.

We used parameters  $T=4.14\MeV$, $V= (18\fm)^3$, $\Delta t=0.49\fm$  and $\Delta x = 1.38\fm$ in our ``best" calculations. The effects due to finite spatial and temporal lattice spacings, as well as finite volume, were controlled by performing calculations at different values of $V$, $\Delta t$ and $\Delta x$. For instance, in order to control for finite volume effects, we performed calculations at three different volumes, $V=(10\fm)^3$, $(14\fm)^3$, and $(18\fm)^3$, while holding $T=4.14\MeV$, $z=1.0$, $\Delta x=2.0\fm$ fixed.  At the largest volumes, errors due to finite volume effects 
are smaller than 
 $2.0 \%$ for $S_V$ and $1.0 \%$ for $S_A$; these we take as  upper bounds on finite volume errors. Similarly, to control for finite $\Delta t$ errors, we performed calculations at four different temporal lattice spacings, $\Delta t=0.49\fm$, $0.31\fm$, $0.245\fm$ and $0.196\fm$, with $T=4.14\MeV$, $V= (6.9\fm)^3$, $\Delta x = 1.38\fm$ held fixed, then extrapolated to the $\Delta t\rightarrow 0$ limit. The typical difference in observables between the $\Delta t\rightarrow 0$ extrapolation and the parameters used in our ``best" simulations is $2\%$ for $S_V$ and $0.5\%$ for $S_A$. 

The extrapolation to the spatial continuum limit ($\Delta x \rightarrow 0$) is the source of the largest systematic errors for most observables
\footnote{This can be seen by considering the Symanzik action~\cite{Symanzik:1983dc,Symanzik:1983gh} (an effective action valid at distances larger than the lattice spacing). The lowest dimension term of this theory not included in~\eq{eq:hamiltonian} involves {\it two} extra derivatives and its coefficient is proportional to $\Delta x^2$ \cite{Nicholson_2017}.}.
To extrapolate to the spatial continuum limit, we perform calculations with three different spatial lattice spacings, $\Delta x = 2.00\fm$, $1.63\fm$ and $1.38\fm$ with  $N_x = 9, 11$ and $13$, while $N_t=96$, $\Delta t = 0.49\fm$ and $z$ are held fixed. We then fit observables to the formula  $\av{\mathcal{O}} = a + b \Delta x^2$. As an example, we show in \fig{fig:dx-cont-lim} the extrapolations of $S_{V/A}(q)$ for a  typical value of $q$ ($q/q_{\text{th}}=0.45$, with $q_{\text{th}}\equiv \sqrt{6MT}\approx 153\MeV$).

\begin{figure}[t!]
\includegraphics[width=0.95\columnwidth, trim=0.2cm 0.75cm 2.5cm 2.5cm, clip]{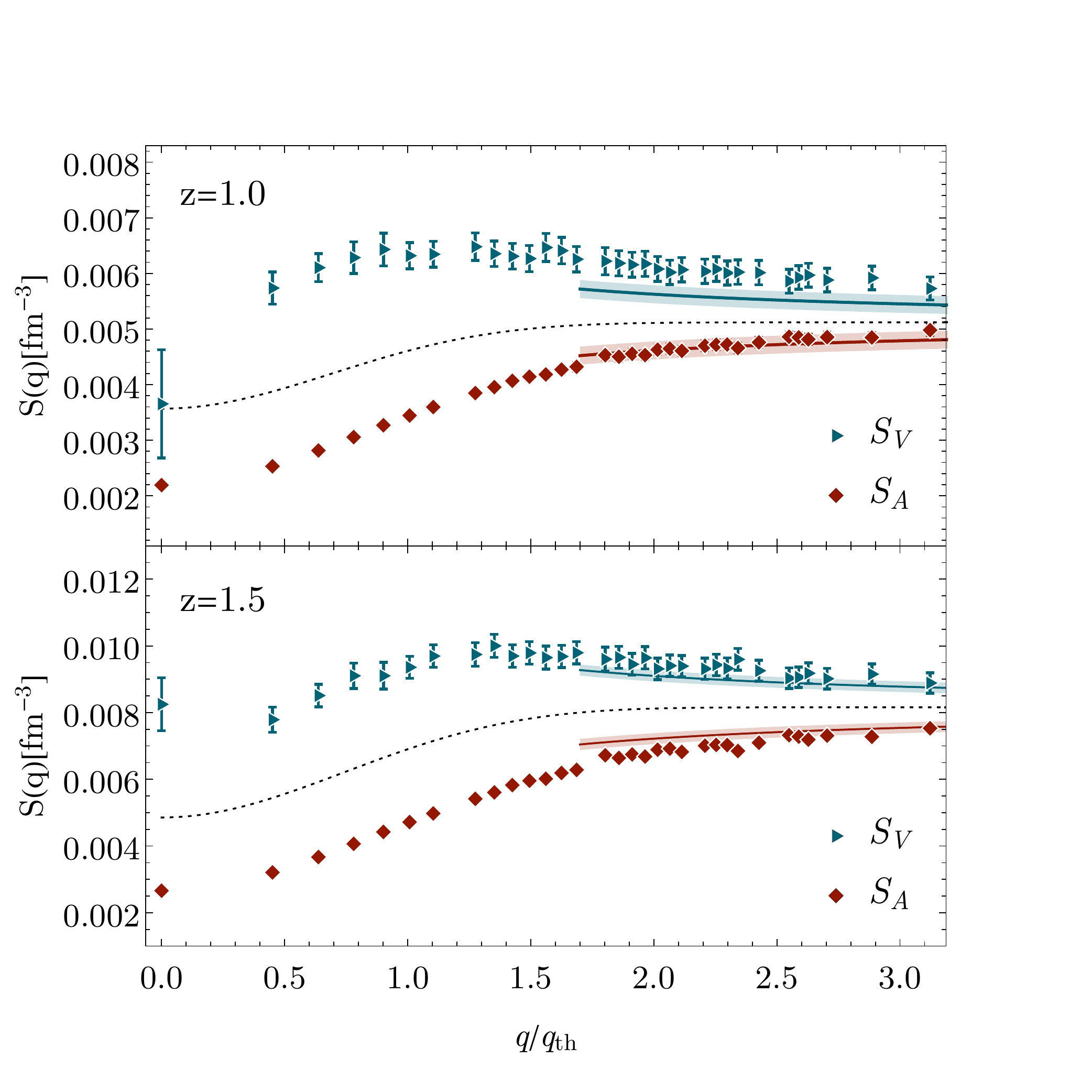}
\caption{Continuum limits of the vector and axial structure factors at fixed temperature and $z=1.0$ and $z=1.5$.
The bands correspond to the OPE asymptotic limits in~\eq{eq:asymptotic} and the dotted lines to the free theory result. 
}
\label{fig:structure-factors}
\end{figure}

Continuum limits for the vector and axial structure factors at fugacities $z=1.0$ and $1.5$ are plotted in \fig{fig:structure-factors}. All sources off error are included in the error bar. This includes statistical errors, the errors due to the $\Delta x \rightarrow 0$ extrapolation (as shown in \fig{fig:dx-cont-lim}), as well as the estimates of the systematic errors due to finite volumes and $\Delta t$ discussed above. Several features of the structure factors deserve comment. First, we find around a $20\%$ suppression in both structure factors relative to the unitary gas when $a=-18.9\fm$ \cite{PhysRevC.101.045805}.\footnote{Similar reductions in the vector structure factor due to a negative scattering length were measured in cold atom experiments \cite{PhysRevLett.105.070402}.} The finite value of the scattering length therefore produces a detectable effect on structure factors. Second, the suppression of the vector structure factor at low momenta is not captured by second order virial calculations~\cite{Bedaque_2018}. On the other hand, fourth order virial calculations of $S_V(q=0)$ and $S_A(q=0)$ for the unitary gas produce qualitatively similar behavior to the $z=1.0$ results of~\fig{fig:structure-factors}~\cite{PhysRevC.96.055804}. However, naively extrapolating these fourth order predictions to $z=1.5$ produces an $S_V(q=0)$ a factor two smaller than \fig{fig:structure-factors} predicts. Therefore, though it is possible that $z=1.5$ may lie within the radius of convergence of the virial expansion, the structure factors computed here cannot be captured by currently available virial coefficients. 

Since the structure factors are derived directly from a partition function they automatically satisfy the following ``sum-rules"
\beq
 S_V(0)  = T \partial n / \partial \mu,  \qquad
  S_A(0)  = T \partial s / \partial h ,\nonumber \\ 
\eeq
required for any thermodynamically consistent theory. This consistency ensures that macroscopic conservation laws are obeyed by these response functions~\cite{PhysRev.124.287}, a feature needed for large-scale supernova simulations~\cite{janka2012corecollapse,Burrows:2018aa}. 
%

The Tan contact parameter, $C$~\cite{TAN20082952}:
\beq
C = \lim_{k\rightarrow \infty} k^4 n(k)~.
\label{eq:contact}
\eeq
characterizes the high momentum behavior of many observables in this system, including both the density and structure factors:
\beqs\label{eq:asymptotic}
S_{V}(q) &= \av{n} + \frac{C}{8 V q} + \mathcal{O}(q^{-2})\\
S_{A}(q) &= \av{n} - \frac{C}{8 V q}  + \mathcal{O}(q^{-2}).
\eeqs
In Ref.~\cite{Braaten_2008} it was shown that this expansion, an example of the operator product expansion (OPE), leads to the relation
\beq\label{eq:ope}
C = g^2 M^2 \int{d^3 x  \langle \psi_1^{\dagger} \psi_1 \psi_2^{\dagger}  \psi_2(x)\rangle}.
\eeq We compute $C$ using \eq{eq:ope} and its continuum extrapolation,  shown in \fig{fig:dx-cont-lim}, gives $C/N k_F=2.1(1)$ (at $z=1.5$) and $C/N k_F=2.1(1)$ (at $z=1.0$). The uncertainty is dominated by statistical and continuum extrapolation errors. In~\fig{fig:structure-factors} we show, besides the continuum extrapolated results for the structure factors, the asymptotic limits at high $q$ predicted by \eq{eq:asymptotic} and our measured values of $\langle n\rangle$ and $C$. The agreement between the OPE prediction \eq{eq:asymptotic} and high-momentum tails of $S_{V/A}$ is a further consistency check for our calculation.

\begin{figure}[t!]
\includegraphics[width=0.95\columnwidth, trim=0.2cm 0.1cm 0cm -0.2cm, clip]{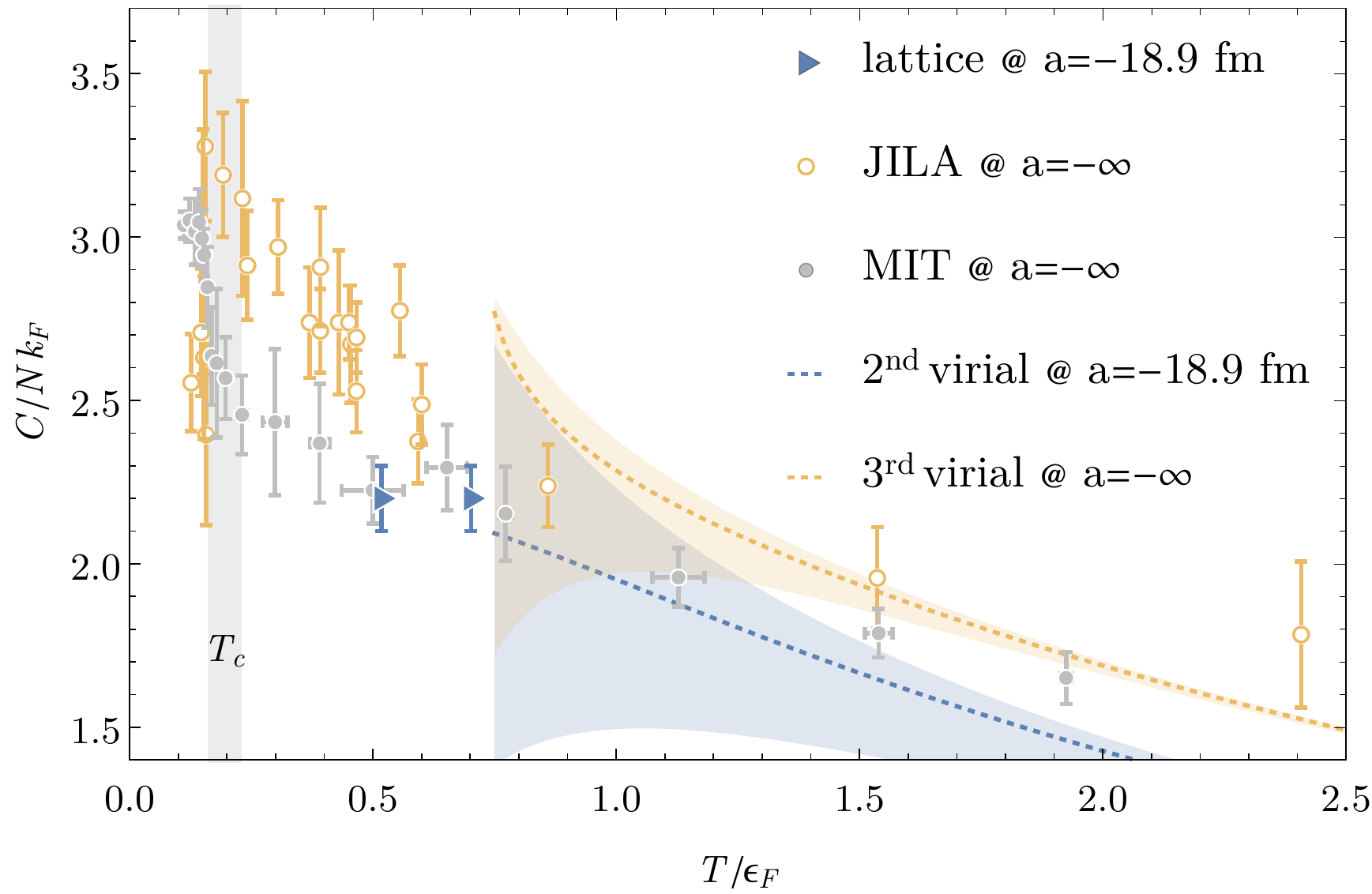}
	\caption{A comparison between the contact at $a=-18.9$ fm (blue points) and the experimentally measured values for the unitary gas (yellow \cite{PhysRevLett.109.220402} and gray points \cite{PhysRevLett.122.203402}). The yellow  curve is the $3^{rd}$ order virial expansion for the unitary gas  \cite{Hu_2011}. The blue line shows the $2^{nd}$ order results for $a=-18.9$ fm. The band around the virial curves incorporates an estimate of the next order contribution.}
\label{fig:contact-plot}
\end{figure}

It is interesting to consider our results for the contact in a broader context. In \fig{fig:contact-plot} we plot our results together with the second order virial expansion prediction. We also plot results obtained with the unitary gas, both experimental and the third order virial prediction. The error bands correspond to estimates of the first neglected order; in the unitary case, the virial coefficients of \cite{Hu_2011,PhysRevLett.116.230401,Nascimb_ne_2010,Ku_2012} were used; at finite scattering length, the formulae of \cite{Hu_2011} we used.

\section{conclusions}

We report on a Monte Carlo calculation of the vector and axial static structure factors of neutron matter in the regime relevant for the physics of supernovae and neutron star mergers.
The hamiltonian describing the strong interactions, leading order pionless effective field theory, is simple, and refinements are both welcome and possible.
All sources of error, statistical and systematic, are accounted for and add up to a few percent.
This control was possible in large part due to technologies seldom used in this context.
The results show a definite change from both the free theory and the unitary limit.
We also calculated the contact, a parameter describing the high momentum distribution of particles, and verified its consistency with the high-momentum dependence of the structure factors.
This calculation opens up a path for a definitive calculation, including a more realistic description of nuclear forces, encompassing most temperatures and densities parameters relevant to supernova physics.

\section{Acknowledgments}
We are grateful to Martin Zwierlien and Biswaroop Mukherjee for providing experimental data. A.A. is supported by U.S. DOE Grant No. DE-FG02-95ER40907. P.B and E.B are supported in part by the US DoE under contract No. DE-FG02-93ER-40762. N.C.W is supported in part by the U.S. DOE under Grant No. DE-FG02-00ER41132.

\bibliographystyle{apsrev4-1}
\bibliography{NeutronsBibliography}

\begin{thebibliography}{27}%
\makeatletter
\providecommand \@ifxundefined [1]{%
 \@ifx{#1\undefined}
}%
\providecommand \@ifnum [1]{%
 \ifnum #1\expandafter \@firstoftwo
 \else \expandafter \@secondoftwo
 \fi
}%
\providecommand \@ifx [1]{%
 \ifx #1\expandafter \@firstoftwo
 \else \expandafter \@secondoftwo
 \fi
}%
\providecommand \natexlab [1]{#1}%
\providecommand \enquote  [1]{``#1''}%
\providecommand \bibnamefont  [1]{#1}%
\providecommand \bibfnamefont [1]{#1}%
\providecommand \citenamefont [1]{#1}%
\providecommand \href@noop [0]{\@secondoftwo}%
\providecommand \href [0]{\begingroup \@sanitize@url \@href}%
\providecommand \@href[1]{\@@startlink{#1}\@@href}%
\providecommand \@@href[1]{\endgroup#1\@@endlink}%
\providecommand \@sanitize@url [0]{\catcode `\\12\catcode `\$12\catcode
  `\&12\catcode `\#12\catcode `\^12\catcode `\_12\catcode `\%12\relax}%
\providecommand \@@startlink[1]{}%
\providecommand \@@endlink[0]{}%
\providecommand \url  [0]{\begingroup\@sanitize@url \@url }%
\providecommand \@url [1]{\endgroup\@href {#1}{\urlprefix }}%
\providecommand \urlprefix  [0]{URL }%
\providecommand \Eprint [0]{\href }%
\providecommand \doibase [0]{http://dx.doi.org/}%
\providecommand \selectlanguage [0]{\@gobble}%
\providecommand \bibinfo  [0]{\@secondoftwo}%
\providecommand \bibfield  [0]{\@secondoftwo}%
\providecommand \translation [1]{[#1]}%
\providecommand \BibitemOpen [0]{}%
\providecommand \bibitemStop [0]{}%
\providecommand \bibitemNoStop [0]{.\EOS\space}%
\providecommand \EOS [0]{\spacefactor3000\relax}%
\providecommand \BibitemShut  [1]{\csname bibitem#1\endcsname}%
\let\auto@bib@innerbib\@empty
\bibitem [{\citenamefont {Burrows}\ \emph {et~al.}(2006)\citenamefont
  {Burrows}, \citenamefont {Reddy},\ and\ \citenamefont
  {Thompson}}]{Burrows:2004vq}%
  \BibitemOpen
  \bibfield  {author} {\bibinfo {author} {\bibfnamefont {A.}~\bibnamefont
  {Burrows}}, \bibinfo {author} {\bibfnamefont {S.}~\bibnamefont {Reddy}}, \
  and\ \bibinfo {author} {\bibfnamefont {T.~A.}\ \bibnamefont {Thompson}},\
  }\href {\doibase 10.1016/j.nuclphysa.2004.06.012} {\bibfield  {journal}
  {\bibinfo  {journal} {Nucl. Phys. A}\ }\textbf {\bibinfo {volume} {777}},\
  \bibinfo {pages} {356} (\bibinfo {year} {2006})},\ \Eprint
  {http://arxiv.org/abs/astro-ph/0404432} {arXiv:astro-ph/0404432} \BibitemShut
  {NoStop}%
\bibitem [{\citenamefont {Bedaque}\ and\ \citenamefont {van
  Kolck}(2002)}]{doi:10.1146/annurev.nucl.52.050102.090637}%
  \BibitemOpen
  \bibfield  {author} {\bibinfo {author} {\bibfnamefont {P.~F.}\ \bibnamefont
  {Bedaque}}\ and\ \bibinfo {author} {\bibfnamefont {U.}~\bibnamefont {van
  Kolck}},\ }\href {\doibase 10.1146/annurev.nucl.52.050102.090637} {\bibfield
  {journal} {\bibinfo  {journal} {Annual Review of Nuclear and Particle
  Science}\ }\textbf {\bibinfo {volume} {52}},\ \bibinfo {pages} {339}
  (\bibinfo {year} {2002})},\ \Eprint
  {http://arxiv.org/abs/https://doi.org/10.1146/annurev.nucl.52.050102.090637}
  {https://doi.org/10.1146/annurev.nucl.52.050102.090637} \BibitemShut
  {NoStop}%
\bibitem [{\citenamefont {Brower}\ \emph {et~al.}(1997)\citenamefont {Brower},
  \citenamefont {Ivanenko}, \citenamefont {Levi},\ and\ \citenamefont
  {Orginos}}]{BROWER1997353}%
  \BibitemOpen
  \bibfield  {author} {\bibinfo {author} {\bibfnamefont {R.}~\bibnamefont
  {Brower}}, \bibinfo {author} {\bibfnamefont {T.}~\bibnamefont {Ivanenko}},
  \bibinfo {author} {\bibfnamefont {A.}~\bibnamefont {Levi}}, \ and\ \bibinfo
  {author} {\bibfnamefont {K.}~\bibnamefont {Orginos}},\ }\href {\doibase
  https://doi.org/10.1016/S0550-3213(96)00579-2} {\bibfield  {journal}
  {\bibinfo  {journal} {Nuclear Physics B}\ }\textbf {\bibinfo {volume}
  {484}},\ \bibinfo {pages} {353 } (\bibinfo {year} {1997})}\BibitemShut
  {NoStop}%
\bibitem [{\citenamefont {Horowitz}\ and\ \citenamefont
  {Schwenk}(2006)}]{HOROWITZ2006326}%
  \BibitemOpen
  \bibfield  {author} {\bibinfo {author} {\bibfnamefont {C.}~\bibnamefont
  {Horowitz}}\ and\ \bibinfo {author} {\bibfnamefont {A.}~\bibnamefont
  {Schwenk}},\ }\href {\doibase https://doi.org/10.1016/j.physletb.2006.09.042}
  {\bibfield  {journal} {\bibinfo  {journal} {Physics Letters B}\ }\textbf
  {\bibinfo {volume} {642}},\ \bibinfo {pages} {326 } (\bibinfo {year}
  {2006})}\BibitemShut {NoStop}%
\bibitem [{\citenamefont {van Kolck}(1999)}]{vanKolck:1998bw}%
  \BibitemOpen
  \bibfield  {author} {\bibinfo {author} {\bibfnamefont {U.}~\bibnamefont {van
  Kolck}},\ }\href {\doibase 10.1016/S0375-9474(98)00612-5} {\bibfield
  {journal} {\bibinfo  {journal} {Nucl. Phys. A}\ }\textbf {\bibinfo {volume}
  {645}},\ \bibinfo {pages} {273} (\bibinfo {year} {1999})},\ \Eprint
  {http://arxiv.org/abs/nucl-th/9808007} {arXiv:nucl-th/9808007} \BibitemShut
  {NoStop}%
\bibitem [{\citenamefont {Kaplan}\ \emph {et~al.}(1998)\citenamefont {Kaplan},
  \citenamefont {Savage},\ and\ \citenamefont {Wise}}]{Kaplan:1998tg}%
  \BibitemOpen
  \bibfield  {author} {\bibinfo {author} {\bibfnamefont {D.~B.}\ \bibnamefont
  {Kaplan}}, \bibinfo {author} {\bibfnamefont {M.~J.}\ \bibnamefont {Savage}},
  \ and\ \bibinfo {author} {\bibfnamefont {M.~B.}\ \bibnamefont {Wise}},\
  }\href {\doibase 10.1016/S0370-2693(98)00210-X} {\bibfield  {journal}
  {\bibinfo  {journal} {Phys. Lett. B}\ }\textbf {\bibinfo {volume} {424}},\
  \bibinfo {pages} {390} (\bibinfo {year} {1998})},\ \Eprint
  {http://arxiv.org/abs/nucl-th/9801034} {arXiv:nucl-th/9801034} \BibitemShut
  {NoStop}%
\bibitem [{\citenamefont {Alexandru}\ \emph
  {et~al.}(2020{\natexlab{a}})\citenamefont {Alexandru}, \citenamefont
  {Bedaque},\ and\ \citenamefont {Warrington}}]{PhysRevC.101.045805}%
  \BibitemOpen
  \bibfield  {author} {\bibinfo {author} {\bibfnamefont {A.}~\bibnamefont
  {Alexandru}}, \bibinfo {author} {\bibfnamefont {P.~F.}\ \bibnamefont
  {Bedaque}}, \ and\ \bibinfo {author} {\bibfnamefont {N.~C.}\ \bibnamefont
  {Warrington}},\ }\href {\doibase 10.1103/PhysRevC.101.045805} {\bibfield
  {journal} {\bibinfo  {journal} {Phys. Rev. C}\ }\textbf {\bibinfo {volume}
  {101}},\ \bibinfo {pages} {045805} (\bibinfo {year}
  {2020}{\natexlab{a}})}\BibitemShut {NoStop}%
\bibitem [{\citenamefont {Duane}\ \emph {et~al.}(1987)\citenamefont {Duane},
  \citenamefont {Kennedy}, \citenamefont {Pendleton},\ and\ \citenamefont
  {Roweth}}]{Duane:1987de}%
  \BibitemOpen
  \bibfield  {author} {\bibinfo {author} {\bibfnamefont {S.}~\bibnamefont
  {Duane}}, \bibinfo {author} {\bibfnamefont {A.}~\bibnamefont {Kennedy}},
  \bibinfo {author} {\bibfnamefont {B.}~\bibnamefont {Pendleton}}, \ and\
  \bibinfo {author} {\bibfnamefont {D.}~\bibnamefont {Roweth}},\ }\href@noop {}
  {\bibfield  {journal} {\bibinfo  {journal} {Phys.Lett.}\ }\textbf {\bibinfo
  {volume} {B195}},\ \bibinfo {pages} {216} (\bibinfo {year}
  {1987})}\BibitemShut {NoStop}%
\bibitem [{\citenamefont {Alexandru}\ \emph
  {et~al.}(2020{\natexlab{b}})\citenamefont {Alexandru}, \citenamefont
  {Bedaque},\ and\ \citenamefont {Warrington}}]{Alexandru:2019gmp}%
  \BibitemOpen
  \bibfield  {author} {\bibinfo {author} {\bibfnamefont {A.}~\bibnamefont
  {Alexandru}}, \bibinfo {author} {\bibfnamefont {P.~F.}\ \bibnamefont
  {Bedaque}}, \ and\ \bibinfo {author} {\bibfnamefont {N.~C.}\ \bibnamefont
  {Warrington}},\ }\href {\doibase 10.1103/PhysRevC.101.045805} {\bibfield
  {journal} {\bibinfo  {journal} {Phys. Rev. C}\ }\textbf {\bibinfo {volume}
  {101}},\ \bibinfo {pages} {045805} (\bibinfo {year} {2020}{\natexlab{b}})},\
  \Eprint {http://arxiv.org/abs/1907.03914} {arXiv:1907.03914 [nucl-th]}
  \BibitemShut {NoStop}%
\bibitem [{\citenamefont {Beth}\ and\ \citenamefont
  {Uhlenbeck}(1937)}]{BETH1937915}%
  \BibitemOpen
  \bibfield  {author} {\bibinfo {author} {\bibfnamefont {E.}~\bibnamefont
  {Beth}}\ and\ \bibinfo {author} {\bibfnamefont {G.~E.}\ \bibnamefont
  {Uhlenbeck}},\ }\href {\doibase
  https://doi.org/10.1016/S0031-8914(37)80189-5} {\bibfield  {journal}
  {\bibinfo  {journal} {Physica}\ }\textbf {\bibinfo {volume} {4}},\ \bibinfo
  {pages} {915 } (\bibinfo {year} {1937})}\BibitemShut {NoStop}%
\bibitem [{\citenamefont {Symanzik}(1983{\natexlab{a}})}]{Symanzik:1983dc}%
  \BibitemOpen
  \bibfield  {author} {\bibinfo {author} {\bibfnamefont {K.}~\bibnamefont
  {Symanzik}},\ }\href {\doibase 10.1016/0550-3213(83)90468-6} {\bibfield
  {journal} {\bibinfo  {journal} {Nucl. Phys. B}\ }\textbf {\bibinfo {volume}
  {226}},\ \bibinfo {pages} {187} (\bibinfo {year}
  {1983}{\natexlab{a}})}\BibitemShut {NoStop}%
\bibitem [{\citenamefont {Symanzik}(1983{\natexlab{b}})}]{Symanzik:1983gh}%
  \BibitemOpen
  \bibfield  {author} {\bibinfo {author} {\bibfnamefont {K.}~\bibnamefont
  {Symanzik}},\ }\href {\doibase 10.1016/0550-3213(83)90469-8} {\bibfield
  {journal} {\bibinfo  {journal} {Nucl. Phys. B}\ }\textbf {\bibinfo {volume}
  {226}},\ \bibinfo {pages} {205} (\bibinfo {year}
  {1983}{\natexlab{b}})}\BibitemShut {NoStop}%
\bibitem [{\citenamefont {Nicholson}(2017)}]{Nicholson_2017}%
  \BibitemOpen
  \bibfield  {author} {\bibinfo {author} {\bibfnamefont {A.}~\bibnamefont
  {Nicholson}},\ }\href {\doibase 10.1007/978-3-319-53336-0_5} {\bibfield
  {journal} {\bibinfo  {journal} {Lecture Notes in Physics}\ ,\ \bibinfo
  {pages} {155}} (\bibinfo {year} {2017})}\BibitemShut {NoStop}%
\bibitem [{\citenamefont {Kuhnle}\ \emph {et~al.}(2010)\citenamefont {Kuhnle},
  \citenamefont {Hu}, \citenamefont {Liu}, \citenamefont {Dyke}, \citenamefont
  {Mark}, \citenamefont {Drummond}, \citenamefont {Hannaford},\ and\
  \citenamefont {Vale}}]{PhysRevLett.105.070402}%
  \BibitemOpen
  \bibfield  {author} {\bibinfo {author} {\bibfnamefont {E.~D.}\ \bibnamefont
  {Kuhnle}}, \bibinfo {author} {\bibfnamefont {H.}~\bibnamefont {Hu}}, \bibinfo
  {author} {\bibfnamefont {X.-J.}\ \bibnamefont {Liu}}, \bibinfo {author}
  {\bibfnamefont {P.}~\bibnamefont {Dyke}}, \bibinfo {author} {\bibfnamefont
  {M.}~\bibnamefont {Mark}}, \bibinfo {author} {\bibfnamefont {P.~D.}\
  \bibnamefont {Drummond}}, \bibinfo {author} {\bibfnamefont {P.}~\bibnamefont
  {Hannaford}}, \ and\ \bibinfo {author} {\bibfnamefont {C.~J.}\ \bibnamefont
  {Vale}},\ }\href {\doibase 10.1103/PhysRevLett.105.070402} {\bibfield
  {journal} {\bibinfo  {journal} {Phys. Rev. Lett.}\ }\textbf {\bibinfo
  {volume} {105}},\ \bibinfo {pages} {070402} (\bibinfo {year}
  {2010})}\BibitemShut {NoStop}%
\bibitem [{\citenamefont {Bedaque}\ \emph {et~al.}(2018)\citenamefont
  {Bedaque}, \citenamefont {Reddy}, \citenamefont {Sen},\ and\ \citenamefont
  {Warrington}}]{Bedaque_2018}%
  \BibitemOpen
  \bibfield  {author} {\bibinfo {author} {\bibfnamefont {P.~F.}\ \bibnamefont
  {Bedaque}}, \bibinfo {author} {\bibfnamefont {S.}~\bibnamefont {Reddy}},
  \bibinfo {author} {\bibfnamefont {S.}~\bibnamefont {Sen}}, \ and\ \bibinfo
  {author} {\bibfnamefont {N.~C.}\ \bibnamefont {Warrington}},\ }\href
  {\doibase 10.1103/physrevc.98.015802} {\bibfield  {journal} {\bibinfo
  {journal} {Physical Review C}\ }\textbf {\bibinfo {volume} {98}} (\bibinfo
  {year} {2018}),\ 10.1103/physrevc.98.015802}\BibitemShut {NoStop}%
\bibitem [{\citenamefont {Lin}\ and\ \citenamefont
  {Horowitz}(2017)}]{PhysRevC.96.055804}%
  \BibitemOpen
  \bibfield  {author} {\bibinfo {author} {\bibfnamefont {Z.}~\bibnamefont
  {Lin}}\ and\ \bibinfo {author} {\bibfnamefont {C.~J.}\ \bibnamefont
  {Horowitz}},\ }\href {\doibase 10.1103/PhysRevC.96.055804} {\bibfield
  {journal} {\bibinfo  {journal} {Phys. Rev. C}\ }\textbf {\bibinfo {volume}
  {96}},\ \bibinfo {pages} {055804} (\bibinfo {year} {2017})}\BibitemShut
  {NoStop}%
\bibitem [{\citenamefont {Baym}\ and\ \citenamefont
  {Kadanoff}(1961)}]{PhysRev.124.287}%
  \BibitemOpen
  \bibfield  {author} {\bibinfo {author} {\bibfnamefont {G.}~\bibnamefont
  {Baym}}\ and\ \bibinfo {author} {\bibfnamefont {L.~P.}\ \bibnamefont
  {Kadanoff}},\ }\href {\doibase 10.1103/PhysRev.124.287} {\bibfield  {journal}
  {\bibinfo  {journal} {Phys. Rev.}\ }\textbf {\bibinfo {volume} {124}},\
  \bibinfo {pages} {287} (\bibinfo {year} {1961})}\BibitemShut {NoStop}%
\bibitem [{\citenamefont {Janka}\ \emph {et~al.}(2012)\citenamefont {Janka},
  \citenamefont {Hanke}, \citenamefont {Huedepohl}, \citenamefont {Marek},
  \citenamefont {Mueller},\ and\ \citenamefont
  {Obergaulinger}}]{janka2012corecollapse}%
  \BibitemOpen
  \bibfield  {author} {\bibinfo {author} {\bibfnamefont {H.~T.}\ \bibnamefont
  {Janka}}, \bibinfo {author} {\bibfnamefont {F.}~\bibnamefont {Hanke}},
  \bibinfo {author} {\bibfnamefont {L.}~\bibnamefont {Huedepohl}}, \bibinfo
  {author} {\bibfnamefont {A.}~\bibnamefont {Marek}}, \bibinfo {author}
  {\bibfnamefont {B.}~\bibnamefont {Mueller}}, \ and\ \bibinfo {author}
  {\bibfnamefont {M.}~\bibnamefont {Obergaulinger}},\ }\href@noop {} {\enquote
  {\bibinfo {title} {Core-collapse supernovae: Reflections and directions},}\ }
  (\bibinfo {year} {2012}),\ \Eprint {http://arxiv.org/abs/1211.1378}
  {arXiv:1211.1378 [astro-ph.SR]} \BibitemShut {NoStop}%
\bibitem [{\citenamefont {Burrows}\ \emph {et~al.}(2018)\citenamefont
  {Burrows}, \citenamefont {Vartanyan}, \citenamefont {Dolence}, \citenamefont
  {Skinner},\ and\ \citenamefont {Radice}}]{Burrows:2018aa}%
  \BibitemOpen
  \bibfield  {author} {\bibinfo {author} {\bibfnamefont {A.}~\bibnamefont
  {Burrows}}, \bibinfo {author} {\bibfnamefont {D.}~\bibnamefont {Vartanyan}},
  \bibinfo {author} {\bibfnamefont {J.~C.}\ \bibnamefont {Dolence}}, \bibinfo
  {author} {\bibfnamefont {M.~A.}\ \bibnamefont {Skinner}}, \ and\ \bibinfo
  {author} {\bibfnamefont {D.}~\bibnamefont {Radice}},\ }\href {\doibase
  10.1007/s11214-017-0450-9} {\bibfield  {journal} {\bibinfo  {journal} {Space
  Science Reviews}\ }\textbf {\bibinfo {volume} {214}},\ \bibinfo {pages} {33}
  (\bibinfo {year} {2018})}\BibitemShut {NoStop}%
\bibitem [{\citenamefont {Tan}(2008)}]{TAN20082952}%
  \BibitemOpen
  \bibfield  {author} {\bibinfo {author} {\bibfnamefont {S.}~\bibnamefont
  {Tan}},\ }\href {\doibase https://doi.org/10.1016/j.aop.2008.03.004}
  {\bibfield  {journal} {\bibinfo  {journal} {Annals of Physics}\ }\textbf
  {\bibinfo {volume} {323}},\ \bibinfo {pages} {2952 } (\bibinfo {year}
  {2008})}\BibitemShut {NoStop}%
\bibitem [{\citenamefont {Braaten}\ and\ \citenamefont
  {Platter}(2008)}]{Braaten_2008}%
  \BibitemOpen
  \bibfield  {author} {\bibinfo {author} {\bibfnamefont {E.}~\bibnamefont
  {Braaten}}\ and\ \bibinfo {author} {\bibfnamefont {L.}~\bibnamefont
  {Platter}},\ }\href {\doibase 10.1103/physrevlett.100.205301} {\bibfield
  {journal} {\bibinfo  {journal} {Physical Review Letters}\ }\textbf {\bibinfo
  {volume} {100}} (\bibinfo {year} {2008}),\
  10.1103/physrevlett.100.205301}\BibitemShut {NoStop}%
\bibitem [{\citenamefont {Hu}\ \emph {et~al.}(2011)\citenamefont {Hu},
  \citenamefont {Liu},\ and\ \citenamefont {Drummond}}]{Hu_2011}%
  \BibitemOpen
  \bibfield  {author} {\bibinfo {author} {\bibfnamefont {H.}~\bibnamefont
  {Hu}}, \bibinfo {author} {\bibfnamefont {X.-J.}\ \bibnamefont {Liu}}, \ and\
  \bibinfo {author} {\bibfnamefont {P.~D.}\ \bibnamefont {Drummond}},\ }\href
  {\doibase 10.1088/1367-2630/13/3/035007} {\bibfield  {journal} {\bibinfo
  {journal} {New Journal of Physics}\ }\textbf {\bibinfo {volume} {13}},\
  \bibinfo {pages} {035007} (\bibinfo {year} {2011})}\BibitemShut {NoStop}%
\bibitem [{\citenamefont {Yan}\ and\ \citenamefont
  {Blume}(2016)}]{PhysRevLett.116.230401}%
  \BibitemOpen
  \bibfield  {author} {\bibinfo {author} {\bibfnamefont {Y.}~\bibnamefont
  {Yan}}\ and\ \bibinfo {author} {\bibfnamefont {D.}~\bibnamefont {Blume}},\
  }\href {\doibase 10.1103/PhysRevLett.116.230401} {\bibfield  {journal}
  {\bibinfo  {journal} {Phys. Rev. Lett.}\ }\textbf {\bibinfo {volume} {116}},\
  \bibinfo {pages} {230401} (\bibinfo {year} {2016})}\BibitemShut {NoStop}%
\bibitem [{\citenamefont {Nascimb{\`e}ne}\ \emph {et~al.}(2010)\citenamefont
  {Nascimb{\`e}ne}, \citenamefont {Navon}, \citenamefont {Jiang}, \citenamefont
  {Chevy},\ and\ \citenamefont {Salomon}}]{Nascimb_ne_2010}%
  \BibitemOpen
  \bibfield  {author} {\bibinfo {author} {\bibfnamefont {S.}~\bibnamefont
  {Nascimb{\`e}ne}}, \bibinfo {author} {\bibfnamefont {N.}~\bibnamefont
  {Navon}}, \bibinfo {author} {\bibfnamefont {K.~J.}\ \bibnamefont {Jiang}},
  \bibinfo {author} {\bibfnamefont {F.}~\bibnamefont {Chevy}}, \ and\ \bibinfo
  {author} {\bibfnamefont {C.}~\bibnamefont {Salomon}},\ }\href {\doibase
  10.1038/nature08814} {\bibfield  {journal} {\bibinfo  {journal} {Nature}\
  }\textbf {\bibinfo {volume} {463}},\ \bibinfo {pages} {1057} (\bibinfo {year}
  {2010})}\BibitemShut {NoStop}%
\bibitem [{\citenamefont {Ku}\ \emph {et~al.}(2012)\citenamefont {Ku},
  \citenamefont {Sommer}, \citenamefont {Cheuk},\ and\ \citenamefont
  {Zwierlein}}]{Ku_2012}%
  \BibitemOpen
  \bibfield  {author} {\bibinfo {author} {\bibfnamefont {M.~J.~H.}\
  \bibnamefont {Ku}}, \bibinfo {author} {\bibfnamefont {A.~T.}\ \bibnamefont
  {Sommer}}, \bibinfo {author} {\bibfnamefont {L.~W.}\ \bibnamefont {Cheuk}}, \
  and\ \bibinfo {author} {\bibfnamefont {M.~W.}\ \bibnamefont {Zwierlein}},\
  }\href {\doibase 10.1126/science.1214987} {\bibfield  {journal} {\bibinfo
  {journal} {Science}\ }\textbf {\bibinfo {volume} {335}},\ \bibinfo {pages}
  {563} (\bibinfo {year} {2012})}\BibitemShut {NoStop}%
\bibitem [{\citenamefont {Sagi}\ \emph {et~al.}(2012)\citenamefont {Sagi},
  \citenamefont {Drake}, \citenamefont {Paudel},\ and\ \citenamefont
  {Jin}}]{PhysRevLett.109.220402}%
  \BibitemOpen
  \bibfield  {author} {\bibinfo {author} {\bibfnamefont {Y.}~\bibnamefont
  {Sagi}}, \bibinfo {author} {\bibfnamefont {T.~E.}\ \bibnamefont {Drake}},
  \bibinfo {author} {\bibfnamefont {R.}~\bibnamefont {Paudel}}, \ and\ \bibinfo
  {author} {\bibfnamefont {D.~S.}\ \bibnamefont {Jin}},\ }\href {\doibase
  10.1103/PhysRevLett.109.220402} {\bibfield  {journal} {\bibinfo  {journal}
  {Phys. Rev. Lett.}\ }\textbf {\bibinfo {volume} {109}},\ \bibinfo {pages}
  {220402} (\bibinfo {year} {2012})}\BibitemShut {NoStop}%
\bibitem [{\citenamefont {Mukherjee}\ \emph {et~al.}(2019)\citenamefont
  {Mukherjee}, \citenamefont {Patel}, \citenamefont {Yan}, \citenamefont
  {Fletcher}, \citenamefont {Struck},\ and\ \citenamefont
  {Zwierlein}}]{PhysRevLett.122.203402}%
  \BibitemOpen
  \bibfield  {author} {\bibinfo {author} {\bibfnamefont {B.}~\bibnamefont
  {Mukherjee}}, \bibinfo {author} {\bibfnamefont {P.~B.}\ \bibnamefont
  {Patel}}, \bibinfo {author} {\bibfnamefont {Z.}~\bibnamefont {Yan}}, \bibinfo
  {author} {\bibfnamefont {R.~J.}\ \bibnamefont {Fletcher}}, \bibinfo {author}
  {\bibfnamefont {J.}~\bibnamefont {Struck}}, \ and\ \bibinfo {author}
  {\bibfnamefont {M.~W.}\ \bibnamefont {Zwierlein}},\ }\href {\doibase
  10.1103/PhysRevLett.122.203402} {\bibfield  {journal} {\bibinfo  {journal}
  {Phys. Rev. Lett.}\ }\textbf {\bibinfo {volume} {122}},\ \bibinfo {pages}
  {203402} (\bibinfo {year} {2019})}\BibitemShut {NoStop}%
\end{thebibliography}%

\end{document}